# Full Bulk Spin Polarization and Intrinsic Tunnel Barriers at the Surface of Layered Manganites

J.W. Freeland[1], K.E. Gray[2], L. Ozyuzer[2*], P. Berghuis[2], Elvira Badica[2], J. Kavich[1], H. Zheng[2] and J.F. Mitchell[2]

[1]*Advanced Photon Source, Argonne National Laboratory, Argonne, IL 60439*

[2]*Materials Science Division, Argonne National Laboratory, Argonne, IL 60439*

**Abstract**

**Transmission of information using the spin of the electron as well as its charge requires a high degree of spin polarization at surfaces. At surfaces however this degree of polarization can be quenched by competing interactions. Using a combination of surface sensitive x-ray and tunneling probes, we show for the quasi-two-dimensional bilayer manganites that the outermost Mn-O bilayer, alone, is affected: it is a 1-nm thick insulator that exhibits no long-range ferromagnetic order while the next bilayer displays the full spin polarization of the bulk. Such an abrupt localization of the surface effects is due to the two-dimensional nature of the layered manganite while the loss of ferromagnetism is attributed to weakened double exchange in the reconstructed surface bilayer and a resultant antiferromagnetic phase. The creation of a well-defined surface insulator demonstrates the ability to naturally self-assemble two of the most demanding components of an ideal magnetic tunnel junction.**

High spin polarization at surfaces and interfaces is a key component for transport of information using the spin degree of freedom of the electron[1]. Unfortunately, the physics and chemistry of surfaces can reduce this spin polarization through chemical inhomogeneity[2,3], strain or surface reconstruction. Many half-metallic ferromagnetic oxides exhibit very high bulk magnetization, but the spin polarization at surfaces and interfaces declines more rapidly than the bulk magnetization as the Curie temperature,

$T_C$, is approached[4]. Magnetic tunnel junctions[5] as well as other probes of surface polarization[6-10] in perovskite-based manganites indicate the presence of a non-ferromagnetic surface region, often called a "dead layer," of thickness ~5 nm.

We show for the quasi-two-dimensional *bilayer* manganites that the outermost Mn-O bilayer, *alone*, is affected: this 1-nm thick intrinsic nanoskin is an insulator with no long-range ferromagnetic order while the next bilayer displays the full spin polarization of the bulk. This unexpectedly abrupt termination is likely due to the reduced dimensionality of the crystal structure. That is, the electronic and magnetic coupling between the bilayers is markedly weaker than the corresponding *intra*bilayer couplings or the isotropic couplings found in non-layered ferromagnetic oxides.

A series of $La_{2-2x}Sr_{1+2x}Mn_2O_7$ single crystals were prepared by the floating zone method[11] between x=0.36 and x=0.4. Sample preparation involved cleaving crystals in air under ambient conditions just prior to mounting in the ultra-high vacuum measurement chamber. The natural cleavage planes between the bilayers (see Fig. 1) provide a clear advantage and both atomic force microscopy and soft-x-ray rocking curves showed large, flat terraces. Surface sensitive absorption measurements compare well with bulk sensitive measurements indicating the surface is not degraded. These experiments involve data from several subsequent runs with different crystal batches which all demonstrate a reproducible non-ferromagnetic surface layer. In addition, their bulk properties have been studied extensively by x-ray and neutron scattering, transport and magnetization[12]. In particular, these are double-exchange[13,14] ferromagnets, that exhibit symbiotic ferromagnetism (FM) and metallic conductivity below $T_C$ of ~120-130 K. Here we use polarized x-ray absorption and scattering (see Methods section, below) at beamline[15] 4-ID-C of the Advanced Photon Source to study magnetism in the surface layers, whilst Au point-contact data and vacuum-gap tunneling are used to assess the metallic or insulating nature of the surface.

X-ray resonant magnetic scattering[16,17] (XRMS) allows us to map directly the magnetization profile near the surface by subtracting data for the x-ray helicity parallel ($I^+$) or anti-parallel ($I^-$) to the magnetic moment (see Fig. 2). We tuned to the Mn L edge and the magnetic moment was aligned parallel to the crystal surface and the scattering plane with an applied magnetic field of 500 Oe, which was determined by field-dependent XRMS to be sufficient to achieve magnetic saturation. Additional magnetic information comes from simultaneous measurements of the polarization-dependent absorption and we refer to the differences ($I^+-I^-$) in absorption as x-ray magnetic circular dichroism[18] (XMCD), which is also shown in Fig. 2.

A change in the *sign* of the XRMS data with incident angle is clearly seen in Fig. 3a, and we interpret this as an interference of x-rays scattering from the non-ferromagnetic (FM) surface layer. For a magnetic profile extending uniformly to the crystal surface, the sign of the XRMS is strictly independent of the scattering angle and is determined by the sign of the absorption XMCD. Crystals *with a non-FM surface layer* will scatter strongly from both the chemical interface (cleaved surface) and the magnetic interface that is deeper down since the chemical and magnetic scattering factors from the Mn 3d electrons are of a similar magnitude[17]. Interference between these two distinct interfaces is readily observable in the XRMS signal. Thus the data of Fig. 3a unequivocally prove that the chemical and magnetic interfaces are out of registry in our layered manganite.

The details of the magnetic profile are determined by modeling the scattering data. Since x-rays at the Mn L edge have a wavelength of ~2 nm, the scattering from the structure in well described by the magneto-optics formalism[19,20]. By assembling a layered structure matching the unit cell, we used the dielectric tensor, determined from the polarization dependent absorption, to model the scattering as a function of the magnetic profile near the surface. The case of a non-FM surface bilayer provides the only match to the angle dependent XRMS (solid lines in Fig 3a). A comparison to simulations of magnetic profiles ranging from zero to two non-magnetic bilayers (see Fig 3b) clearly demonstrates that a *single bilayer alone* is non-FM. Calculations show that as the second bilayer magnetization decreases, the XRMS peak position shifts continuously between the two positive peaks shown in Fig. 3b. From this we estimate the spin polarization in the second bilayer at 35 K is the same as the bulk value, with an uncertainty <20%. The XRMS data thus reveal the presence of an extremely thin nonmagnetic 'nanoskin,' below which subsequent bilayers are fully magnetized.

The near surface XMCD data of Fig. 2, measured by electron yield, is about half the size of previous data on a 3-D perovskite manganite[6] at low temperature. This finding is consistent with a non-FM surface bilayer since the ratio is averaged over the escape of electrons with a mean free path of ~1.5 nm. We estimate that the integrated intensity from the top 1-nm (bilayer) corresponds to ~50% of the total signal, so this ratio also requires nearly full bulk magnetization in the second bilayer.

In this field geometry, we are predominately sensitive to in-plane moments so if the spins in the outermost bilayer exhibited FM order that was perpendicular to the layers, it would not be readily visible in the above data. However, we dismiss this for the following reasons. First, additional XMCD data were obtained in fields up to 7 T. For perpendicular field we found no net remanent moment. For parallel fields, XMCD shows a low-field saturation of magnetization in the second bilayer plus a linear component from the surface bilayer, which reaches ~20% of saturation at 7 T. Assuming this increase is due to perpendicular FM order in the surface bilayer, this increase would imply a huge uniaxial anisotropy of $>10^8$ erg/cm$^3$. That is near the upper limit of known materials that contain 4f electrons and significant orbital moment. The layered manganites contain 3d electrons and their orbital moment is quenched, resulting in typical[21] anisotropy of $<10^6$ erg/cm$^3$. In addition, the surface bilayer is shown below to be an insulator: this precludes FM double exchange, and insulating layered manganites are antiferromagnetic (AF). Thus it seems reasonable that the enhanced surface magnetization at 7 T results from overcoming the AF superexchange energy (typically ~$10^8$ erg/cm$^3$ in these materials[22]) in the outermost bilayer. To confirm this idea we measured a layered manganite with x=0.5, which is an A-type AF below 200 K that is insulating in the bulk[12]. Here the A-type structure consists of two sheets of FM spins within the bilayer that are oriented antiferromagnetically. While there is no net moment at zero field for x=0.5, a magnetic field cants the FM sheets to produce a moment and the resulting magnetization measured by a magnetometer and XMCD are linear in field and roughly equal (within 10%). The magnetization increase for x=0.5 at 7 T is close to the ~20% found above in the surface bilayer for x=0.36. From this we conclude that the surface bilayer for x=0.36 exhibits AF order, presumably because the double exchange is preferentially weakened by the surface reconstruction.

We now address the temperature dependence of the spin polarization in the second bilayer. All other half-metallic oxides[4] show a significant loss of surface spin polarization as $T_C$ is approached. For example, in the $La_{1-x}Sr_xMnO_3$ system, the 'near surface' shows significant degradation from the bulk magnetization[6] well below $T_C$ and the 'very surface' displays a nearly linear decrease[7] for all T up to $T_C$. Such a strong temperature-dependent degradation is also implied at interfaces in $La_{1-x}Sr_xMnO_3$ magnetic tunnel junctions[5], which show a complete loss of magnetoresistance for T less than ~0.7 $T_C$. In stark contrast, Fig. 3c compares the temperature dependence of the ratios $(I^+-I^-)/(I^++I^-)$ for XMCD and XRMS with the bulk magnetization[23] determined from neutron scattering. For this sample (x=0.36) the second bilayer retains the bulk value of the magnetization to within ~ 10 K $T_C$. The *only* difference for x=0.4, was the deviation from the bulk begins ~30 K below $T_C$.

Now we address the question of whether the non-FM surface bilayer is metallic. The electronic properties of the c-axis faces of these single crystals were probed with both scanning tunneling spectroscopy (STS) and gold-tip point contacts on crystals that were cleaved in air at room temperature. The vacuum gap in STS guarantees tunneling while the gold point contacts can only exhibit tunneling if an insulating barrier is present on the surface of the crystal. The data of Fig. 4 include numerous high-resistance Au point contacts taken sequentially at 4.2 K on crystals with x=0.36 plus representative STS data taken well above $T_C$. They all exhibit a small degree of asymmetry with the larger current for the crystal positive. This sequence at 4.2 K demonstrates the ability of the Au tip to clean the surface of contamination (presumably physisorbed). The initial 'gentle' touches (open circles and squares) yield featureless log current vs. voltage curves, like the high-temperature data (inverted triangles) using STS that cannot clean the surface. Harder contacts significantly deform the Au tip, and in doing so progressively scrub the surface clean. The open-diamond data were taken after two hard contacts and the three solid-symbol data sets show the limiting characteristic after ~10 hard contacts (note that the solid squares are taken on a different x=0.36 crystal surface after many hard contacts). The three solid-symbol data sets are also shown in the linear plot of the inset of Fig. 4.

These three sets of solid data points fit tolerably well, over a range of 10,000 in current, to the calculation of tunneling (thick lines) through a square barrier of height 375 meV using widths of 1.4 nm (solid squares) and 1.5 nm (solid circles and diamonds). We note that bandgaps of ~300 meV have been observed[24] by STS in both the high-temperature paramagnetic and the low-temperature charge-ordered *insulating states* of the non-layered manganite, $Bi_{1-x}Ca_xMnO_3$. We also note that the exposed surface after cleaving the layered manganite crystals should be at the weakly bonded symmetry plane between the bilayers[12]. Thus if the surface bilayer is insulating, the tunneling distance from the topmost Mn-O layer of the second bilayer to the Au point contact would be ~1.4 nm in the absence of surface reconstruction (likely expansion) in the insulating surface bilayer. We also point out that the open-diamond data in Fig. 4 is reproduced by summing parallel contributions from (I) the featureless (surface-contaminated) data of the open squares of Fig. 4 and (II) the tunneling calculation that fits the solid-symbol data. This last observation is consistent with an incompletely cleaned surface after two 'hard' contacts. The distinct, reproducible barrier feature seen in the solid-symbol data of Fig. 4 attests to a high degree of barrier uniformity that may only be possible with an intrinsic barrier. Variations in barrier height, e.g., due to surface contamination, would spread out the feature in energy (i.e., voltage), but the quality of the fit dismisses such variations. Likewise, the highly reproducible current-step *size*, from below to above 375 mV, rules out extrinsic variations in barrier width that would affect the current-step size exponentially.

Unfortunately, obtaining and/or maintaining a clean surface is far more difficult at higher temperatures where the benefits of cryopumping are compromised. Even after repeated hard contacts, data taken at 77 K showed only occasionally a hint of the weak structure seen in the open-diamond data of Fig. 4. At temperatures above $T_C$, all data were featureless, emulating that of the contaminated surfaces taken after 'gentle' touches at 4.2 K. This similarity may imply that the density-of-states for tunneling (through the contamination-layer/insulating-surface-bilayer combination) is insensitive to whether the bulk manganite beneath is a bad metal (below $T_C$) or a bad insulator (above $T_C$). One should recognize that in high fields the conductivity by thermal activation above $T_C$ approaches the low-temperature metallic conductivity (see Fig. 7 of Ref. 25), with its short-mean-free-path[26] of ~1.5 nm. This comparable conductivity for the insulator results from a significant lowering of the thermal activation barrier due to double exchange when high fields align the spins[27]. Since for tunneling to a non-magnetic counterelectrode, e.g., Au, the double-exchange activation barrier is completely absent, we may anticipate little effect except for the possible appearance of a gap at the Fermi level. The present data do not consistently distinguish such a gap, but do not rule it out.

Since the crossover from the metallic state at 4.2 K to the insulating state above $T_C$ was inaccessible, point-contact tunneling data were collected on an insulating layered manganite crystal with x=0.48, which exhibits an insulator-metal transition[25] in fields of 3-4 T (Figs. 5b and 5c) at 4.2 K. The featureless, zero-field tunneling curve shown in Fig. 5a was repeatable even after the requisite number of hard contacts to clean the surface. In zero-field, the high resistance of the insulating crystal also contributes to the point-contact voltage. Upon increasing the field to 6 T, the crystal resistance drops by more than four orders-of-magnitude[25] and the characteristic shoulder seen in the solid data symbols of Fig. 4 appeared, which shows the presence of an insulating surface bilayer. The known magneto-thermal hysteresis in this material[25] is reproduced in the tunneling—the shoulder does not disappear upon lowering the field to zero at 4.2 K, but it disappeared after briefly cycling the crystal in zero field to 95 K and back to 4.2 K.

Thus at least at low temperatures, we conclude that the topmost bilayer is an insulator with a 375-meV bandgap. Therefore, the insulating and non-FM regions coincide. The simultaneous absence of magnetic order and metallic conductivity in the top bilayer is entirely consistent with the double-exchange mechanism[13,14] used to explain conductivity and ferromagnetism below $T_C$ in manganites. The abrupt changes in *only* the topmost bilayer are likely due to the weak electronic and magnetic coupling between bilayers engendered by the crystal structure[12] and not found in non-layered magnetic oxides.

Due to the complex interplay of the structural, electronic, orbital and magnetic degrees of freedom there exist many possibilities for the formation of an insulating, non-FM layer at the surface. While our results do not directly pinpoint the origin of this surface layer, they can eliminate some possibilities. Surface roughness can be ruled out due to our observation of large atomically flat terraces created upon cleaving the crystal. Significant deviations from stoichiometry at the surface due to Sr segregation, which is observed in the perovskite system[3], or oxygen loss would result in a change in the electronic structure at the surface (from the $I^+ + I^-$ data for both XRMS and XMCD) that is much larger that observed. If such disorder is ruled out, a prime candidate may be surface strain[28] due to surface reconstruction or a change in atomic coordination[29] that could stabilize orbital- or charge-ordered insulating states that, in turn, would suppress the FM double-exchange mechanism.

In summary, we have shown that the surface bilayer of air-cleaved layered manganites forms an antiferromagnetic insulating nanoskin comprised of a single bilayer unit, and the next bilayer is metallic and retains the full bulk spin polarization up to nearly $T_C$. While the surface is prepared in air under ambient conditions, indications are that, since no strong bonds are broken during the cleavage process, the surface is stable and the results represent an intrinsic property. An important aspect of this result is that the outermost bilayer could act as an intrinsic tunnel barrier between the fully spin-polarized bilayer beneath and a subsequently deposited FM counter electrode. Artificially grown oxide heterostructures are typically plagued by inhomogeneities in the barrier that can degrade performance. In contrast, for layered manganite crystals, nature provides this tunnel barrier with uniform thickness and properties.

**Methods**

The resonant absorption and scattering processes operate by tuning the x-ray energy to a threshold corresponding to excitation from a 2p core level to the unoccupied 3d states of Mn. Since this energy is specific to Mn, it provides an elemental selective way to study the electronic structure of Mn. If we add polarization dependence then magnetic information can be extracted as well. In practice this involves measurements of the signals with the sample magnetization parallel ($I^+$) or anti-parallel ($I^-$) to the incident beam polarization. With the circularly polarized beam incident at grazing angles of 5 to 16 deg., a constant magnetic field of +/- 500 Oe was then applied in the plane of the surface at each point of the energy scan to achieve this configuration. Field dependent measurements performed in-situ show 500 Oe is sufficient to achieve magnetic saturation for all the samples. The average ($I^+ + I^-$) consists of purely electronic information while the difference ($I^+ - I^-$) is magnetic in origin. For the case of absorption this signal is purely magnetic, but in the case of scattering it consists of a charge-magnetic interference term. It is this term that provides the contrast necessary to extract the magnetic profile.

It is important to recognize that for low-specific resistance junctions, the spreading resistance in the fairly resistive manganite 'substrate' contributes to the measured voltage. Spectral features of such point contacts exhibit a magnetic-field dependence in excellent agreement with the resistivity of the layered manganite crystal measured in a separate experiment by the methods discussed in Ref. 25 and 26. Thus we have only used data from our highest resistance Au point-contact junctions in Fig. 4, although lower resistance junctions were made during the 'hard' contacts used for cleaning the manganite surface.

**Acknowledgements**

The research, including the use of the Advanced Photon Source, was supported by the US Department of Energy (DOE), Basic Energy Sciences, under contract W-31-109-ENG-38.

Correspondence and requests for materials should be addressed to JWF, freeland@anl.gov

**Fig. 1.** Structure of the naturally bilayered manganite, $La_{2-2x}Sr_{1+2x}Mn_2O_7$. The $MnO_6$ octahedra denoted in blue and (La,Sr) sites shown as yellow and red dots. The bilayer repeat distance is ~10 Å. The non-ferromagnetic surface bilayer is colored brown.

**Fig. 2.** Polarization dependent scattering and absorption data at the Mn $L_3$ edge. The sum and difference of signals at 35 K for the x-ray helicity parallel ($I^+$) or anti-parallel ($I^-$) to the magnetic moment for the x=0.36 sample. The XRMS is extremely sensitive to the magnetization profile, while the magnetic properties averaged over the near surface region are probed by XMCD. Electronic properties are measured by $I^++I^-$.

**Fig. 3** Magnetic profile determination and temperature dependence of the bulk like sub-surface bilayer. **a.** The observed sign change of our XRMS data at two angles can only be simulated by chemical and magnetic interfaces that are out of registry. Data were taken at 35 K. **b.** Calculated XRMS energy dependence at $\theta = 16$ degrees for the case of zero, one and two non-ferromagnetic (non-FM) bilayers. From the profound lineshape change we can pinpoint that the data are consistent only with the case of a single non-FM bilayer. **c.** Normalized temperature-dependent magnetization of the second bilayer as measured by the XRMS (circles) and near-surface probe XMCD (squares). These are virtually identical and show slight differences, only very close to $T_C$, with the bulk magnetization from neutron scattering[23] (diamonds).

**Fig. 4.** The log current-voltage characteristics of numerous high-resistance Au point-contact (PT) junctions taken at 4.2 K on a crystal x=0.36 are plotted together with representative STS data taken well above $T_C$. The curves have been displaced for clarity of presentation---the unknown junction areas scale the actual currents. Five of the PT curves were part of a sequence, and they demonstrate the surface cleaning ability of the malleable Au tip when it is deformed by hard contact. The open circles and squares were taken during initial 'soft' contacts. The open diamond data was taken after two 'hard' contacts and the limiting behavior of the solid diamonds and circles were taken after ~10 'hard' contacts. The solid squares represent the limiting case for another cleaved surface on a crystal with x=0.36. The thick lines are fits to the standard tunneling model for the three sets of solid-symbol data using a barrier height of 375 meV and a width of 1.4-1.5 nm. The inability of the scanning tunneling microscope tip to clean the surface lead to the featureless STS curve that is similar to the PT data in the limit of only non-cleaning 'soft' touches. Inset: solid-symbol data on a linear scale.

**Fig. 5.** Field induced metal-insulator transition for a layered manganite with x=0.48. (a) Gold point-contact data showing how the featureless zero-field point-contact curve switches in 6 T to one showing a barrier, similar to that seen in the solid data of Fig. 4, on a metallic bulk manganite crystal. This composition exhibits an insulator-metal transition at 4.2 K in a magnetic field of 3-4 T as seen in (b) for both the c-axis and ab-plane conductivity for x=0.48. This conductivity data is taken from Ref. 25 on a crystal showing very similar magnetization data as that shown in (c) the crystal used for the tunneling of (a).

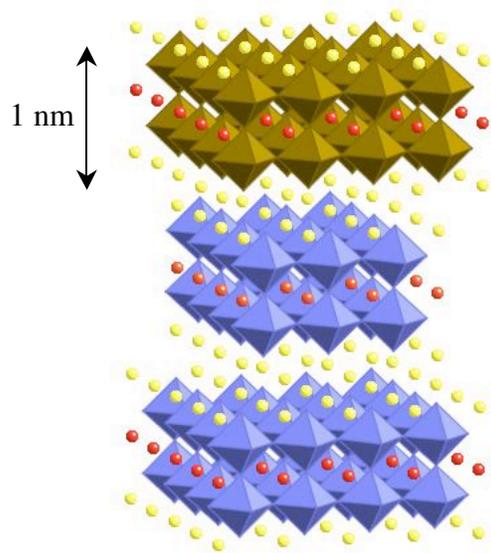

**Fig. 1.**

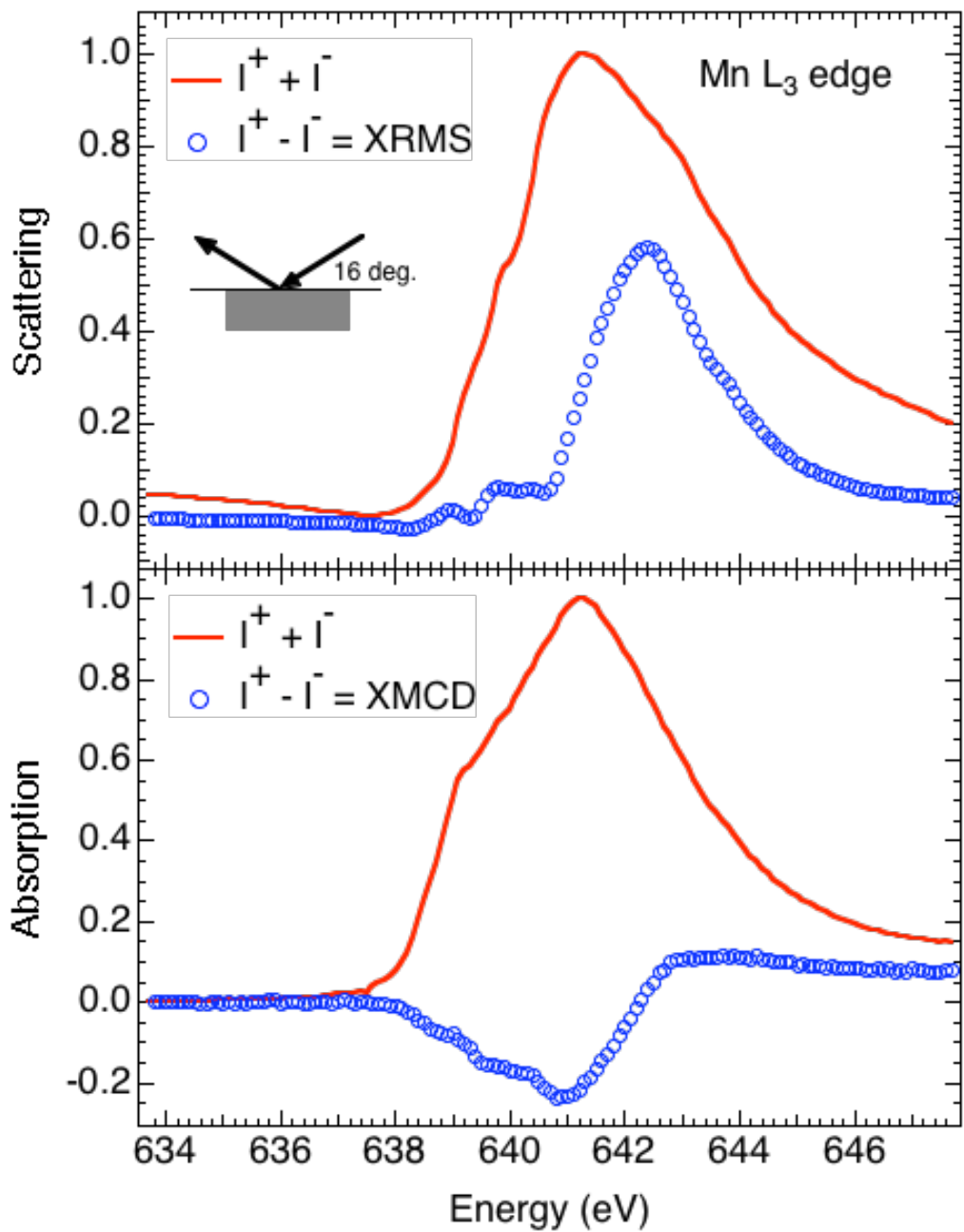

**Fig. 2.**

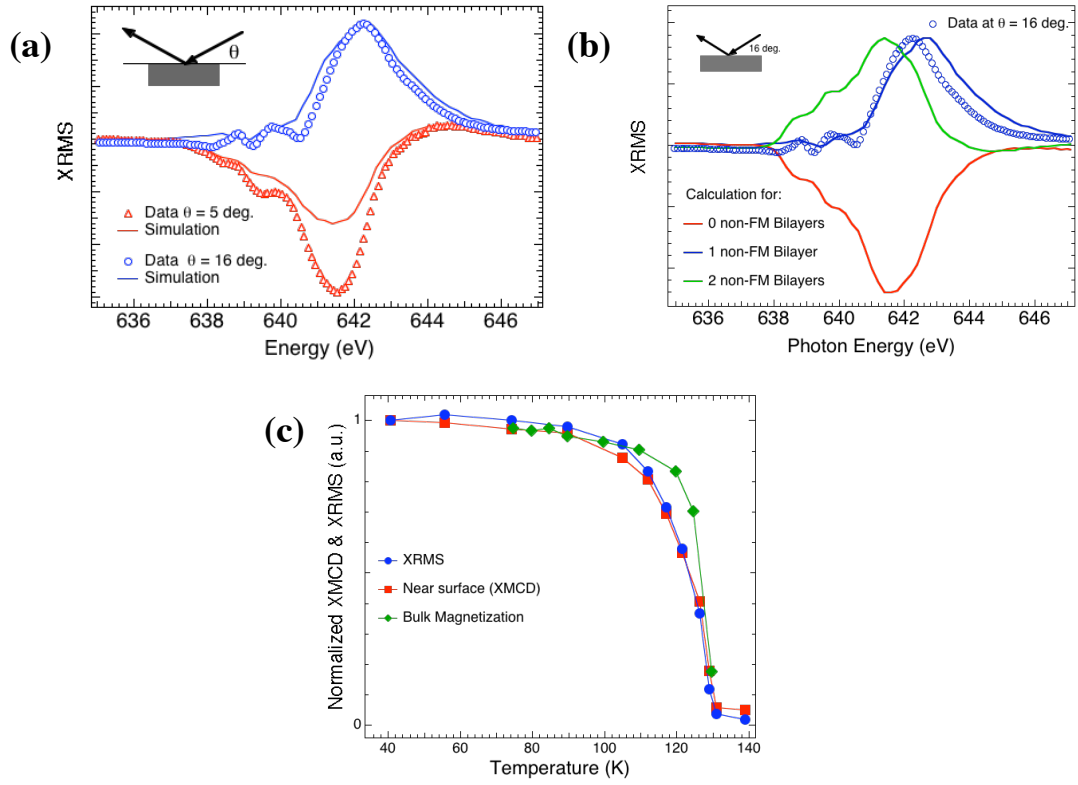

**Fig. 3.**

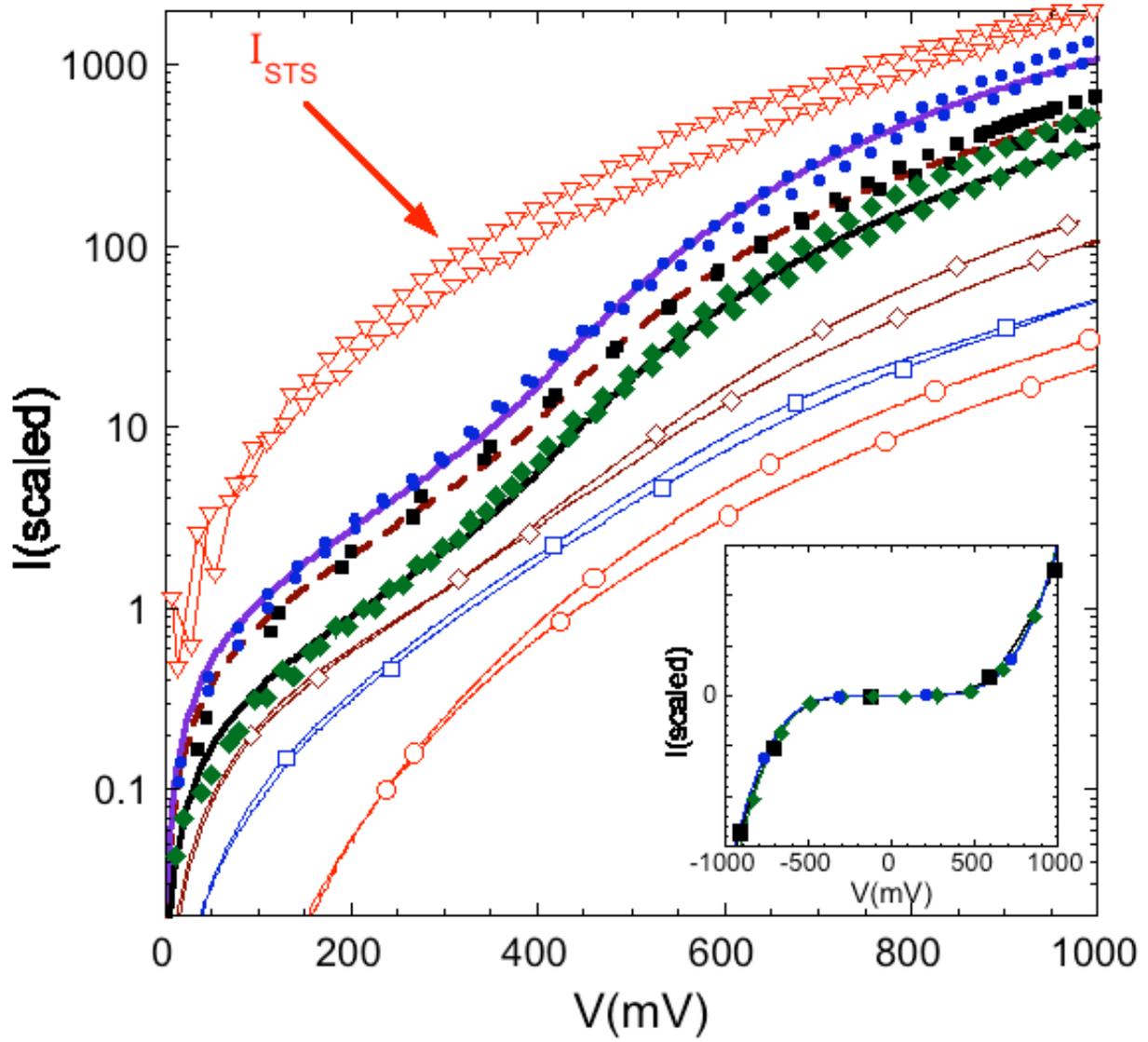

**Fig. 4.**

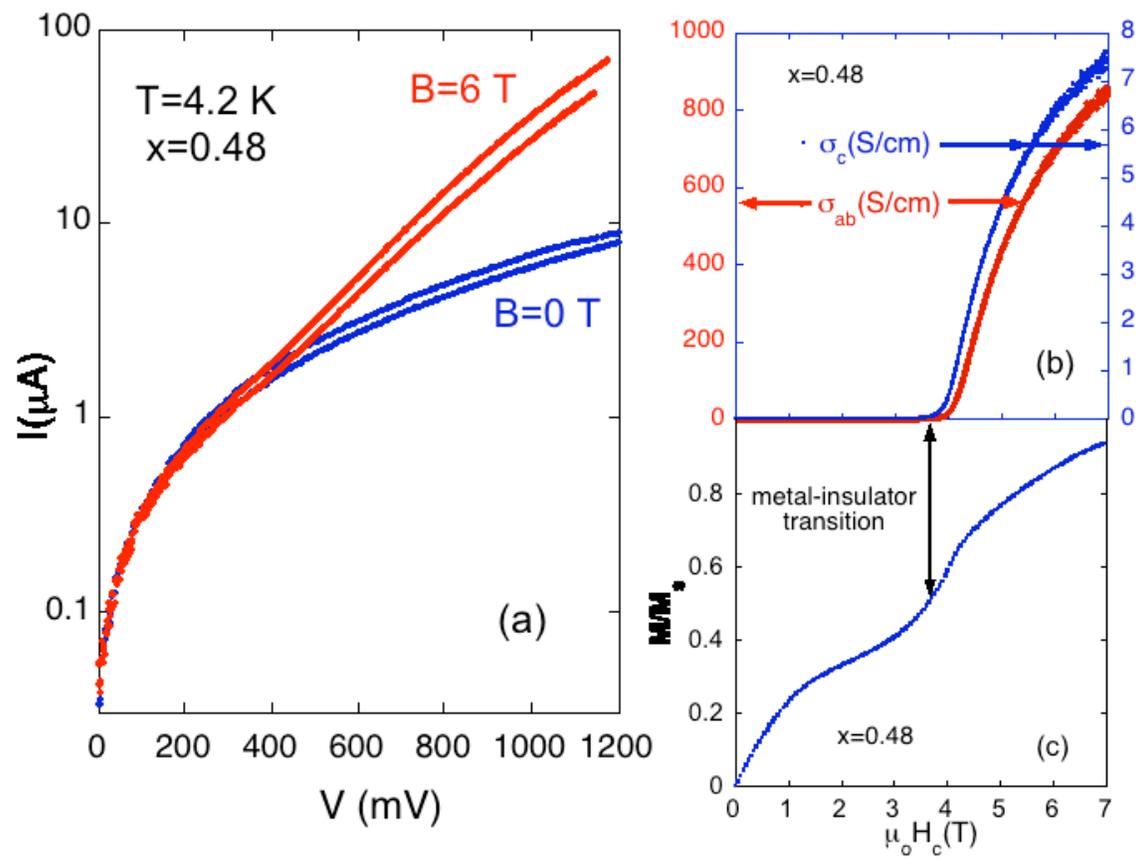

**Fig. 5**